\documentclass[twocolumn,prd,showpacs,preprintnumbers]{revtex4}
\usepackage{amsmath} 
\usepackage{amsfonts} 
\usepackage{amssymb}

\begin{document}

\preprint{KEK-TH-1136}
\preprint{OIQP-07-02}

\title{
Higher-spin Currents and Thermal Flux from Hawking Radiation
} 
\author{Satoshi Iso$^{a}$} 
\email{satoshi.iso@kek.jp}
\author{Takeshi Morita$^a$}
\email{tmorita@post.kek.jp}  
\author{Hiroshi Umetsu$^b$}
\email{hiroshi_umetsu@pref.okayama.jp}

\affiliation{
$^a$ Institute of Particle and Nuclear Studies, High Energy Accelerator
Research Organization(KEK),  
Oho 1-1, Tsukuba, Ibaraki 305-0801, Japan \\
$^b$ Okayama Institute for Quantum Physics,
Kyoyama 1-9-1, Okayama 700-0015, Japan
}
\begin{abstract} 

Quantum fields  near black hole horizons can be
described in terms of an infinite set of $d=2$ conformal fields.  In this
paper, by investigating transformation properties of general higher-spin
currents  under a conformal transformation, we reproduce
the thermal distribution of Hawking radiation in both cases of bosons and
fermions.
As a byproduct, we obtain a generalization
of the Schwarzian derivative for higher-spin
currents.
\end{abstract}
\pacs{04.62.+v, 04.70.Dy, 11.30.-j }

\maketitle


\section{Introduction}

Hawking studied quantum effects of matter in the background of a black hole
formed in collapse and concluded that the black hole will emit thermal
radiation as if it were a blackbody at the Hawking temperature
\cite{Hawking:sw, Hawking:rv}.  Soon after Unruh \cite{Unruh:db} realized
importance of choice of vacua and showed that eternal black holes emit the
same radiation. There are many other derivations and all of them
universally give the same result.

Recently Robinson and Wilczek proposed a new partial derivation of Hawking
radiation \cite{Robinson:2005pd}, which ties its existence to the
cancellation of gravitational anomalies at the horizon.  The method was then
generalized in \cite{IUW1} to charged black holes by using the gauge anomaly
in addition to the gravitational anomaly and further applied to rotating
black holes \cite{IUW2}\cite{Murata} and others
\cite{Das, setare, Xu, IMU, Jiang, Jiang2}.  
The essential observation is that quantum fields near the horizon behave as an
infinite set of two-dimensional fields and ingoing modes at the horizon can
be considered as left moving modes while outgoing modes as right moving
modes.  Once the ingoing modes fall into the black hole, they never come out
classically and cannot affect the physics outside the black hole.  Quantum
mechanically, however, they cannot be neglected because, without the ingoing
modes, the theory becomes chiral at the horizon, which makes the effective
theory anomalous under general coordinate or gauge transformations.  In this
sense, the ingoing modes at the horizon only affect the exterior region
through quantum anomalies.

The derivation of the Hawking flux through quantum anomalies indicates
universality of the Hawking radiation, but it is so far only partial because
the full thermal distribution has not yet been obtained.  Schwarzschild
black hole emits thermal radiation with the Planck distribution $N^\pm
(\omega)$.  Here $N^\pm(\omega)$ is given by
\begin{equation}
N^\pm (\omega) =\frac{1}{e^{\beta \omega} \pm 1},
\end{equation} 
where $1/\beta$ is the Hawking temperature of the black hole
and $\pm$ corresponds to fermions and bosons respectively. 
The energy flux is given by
the following specific moment $F_2^\pm$
of $N^\pm(\omega)$, 
but in order to derive the complete
thermal radiation it is necessary to obtain all the fluxes with higher
moments $F_n^\pm$
\footnote{Since $N^\pm(\omega)$ is defined
for positive $\omega$, it is sufficient to consider
even $n$. }.  
Here we define $F^\pm_n$ ($n \geq 1$) as an $(n-1)$-th
moment of the Planck distribution
\begin{equation}
F_n^{\pm}=\int_0^\infty \frac{d\omega}{2\pi} 
 \omega^{n-1}  N^\pm (\omega).
\end{equation}
The integrals can be performed as
\begin{eqnarray}
 F_{2n}^+ &=&  \left(1 - 2^{1-2n}\right)\frac{B_n}{8\pi n} \kappa^{2n}, 
  \nonumber \\
 F_{2n}^- &=& \frac{1}{8\pi n} B_n \kappa^{2n},  
\end{eqnarray} 
where $B_n$'s are the Bernoulli numbers ($B_1=1/6$, $B_2=1/30$)
and $\kappa=2\pi/\beta$ is the surface gravity of the
black hole.

$F_2^\pm$ is the energy flux from the black hole.  Similarly $F_{2n}^\pm$
can be expected to be given by fluxes of higher-spin currents.  Since each
partial wave of quantum fields near the horizon is described by $d=2$
conformal field, there are infinitely many higher-spin conserved currents
near the horizon.  To determine their fluxes in a similar method adopted for
the energy flux \cite{Robinson:2005pd} or the charge flux \cite{IUW1}, we
need to calculate quantum anomalies for higher-spin currents to fix the
boundary conditions of the currents at the horizon.  Covariant calculations
of these anomalies are quite involved and we leave them for a future
publication \cite{IMU2}.  In this paper, we use a much simpler and powerful
technique of conformal field theories to derive the complete thermal
spectrum of Hawking radiation.

\section{Energy flux}

We first review a derivation of 
the energy flux from Hawking radiation for
Schwarzschild black hole by using a conformal field theory technique
\cite{Thorlacius}\cite{Strm}.  
Since each partial wave of quantum fields in a
Schwarzschild black hole background behaves as an
independent two-dimensional conformal field, we can treat them separately
in the two-dimensional $(t, r)$ plane. 
Covariant energy-momentum tensor 
$T_{\mu\nu}=(2/\sqrt{-g}) \delta S/\delta g^{\mu\nu} $ 
is classically traceless and conserved but when the matter field is 
quantized they give rise to a conformal anomaly
$T^{\mu}_{\mu}=c/24\pi R$ where $c$ is 
the  central charge of the matter
field. In the conformal gauge $g_{\mu \nu}=e^{\varphi} \eta_{\mu \nu}$, the
conservation of the energy momentum tensor can be solved and the $(uu)$
component becomes
\begin{equation}
 T_{uu}(u, v) = \frac{c}{24 \pi} 
  \left(
   \partial_u^2 \varphi -\frac{1}{2}(\partial_u \varphi)^2 
  \right)
  + T_{uu}^{(conf)}(u) .
  \label{emtensor}
\end{equation} 
Here $u=t-x$ and $v=t+x$ are light cone coordinates.  
$T_{uu}^{(conf)}(u)$ is  holomorphic (i.e. independent of  $v$) 
and independent of the conformal factor $\varphi$.  
We need a boundary condition to determine
$T_{uu}^{(conf)}(u)$ in the black hole background. 
This is essentially the same as 
the derivation given by Christensen and Fulling \cite{CF}.
Under a holomorphic coordinate transformation
 $(u,v) \rightarrow (w(u),y(v))$, since
$T_{uu}$ transforms covariantly and the conformal factors are related as
$\tilde{\varphi}(w,y)=\varphi(u, v) -\ln \frac{dw}{du} \frac{dy}{dv} $, 
the holomorphic part transforms as
\begin{equation}
 T_{ww}^{(conf)}(w)
  = \left(\frac{dw}{du} \right)^{-2} 
  \left(T_{uu}^{(conf)}(u) + 
   \frac{c}{24 \pi} \{w,u \} \right),
\end{equation}
where $\{w, u \}$ is the Schwarzian derivative,
\begin{equation}
\{w,u \} = \frac{w'''}{w'}  -\frac{3}{2} \left(\frac{w''}{w'}\right)^2.
\end{equation}
The prime means a derivative with respect to $u$.

In the case of Schwarzschild black hole, 
we are interested in the outgoing energy flux measured by asymptotic
inertial observers whose natural coordinates are asymptotic Minkowski
variables $(u,v)$ where $u=t-r_*$  
($v=t + r_*$) are outgoing( ingoing) light-cone coordinates, 
and $r_*$ is the tortoise
coordinate.  Regular coordinates near the horizon, on the other hand, are
given by 
the Kruskal coordinates $(U,V)$ which are related to $(u,v)$ as
$U=-e^{-\kappa u}$ and $V=e^{\kappa v}$.  
Hence we have a relation
\begin{equation}
T_{UU}^{(conf)}(U)= \left(\frac{1}{\kappa U}\right)^2   
\left( T_{uu}^{(conf)}(u) + \frac{c}{24 \pi} \{U,u \} \right).
\label{conftr}
\end{equation}
As in \cite{Unruh:db} we impose a boundary condition for the outgoing energy
flux such that physical quantities should be regular at the future horizon
$U=0$ in the Kruskal coordinate.  
This requires that $\langle T_{UU}^{(conf)}(U) \rangle$
is finite at the horizon, and hence $\langle T_{uu}^{(conf)}\rangle$ 
is determined to be $-\left(c/24\pi\right)\{U,u \}$. 
We further assume that there is no ingoing flux at infinity. 
As a result we  obtain the asymptotic flux by the value of
the Schwarzian derivative
\begin{equation}
\langle T^r_t \rangle= \langle T_{uu}\rangle-\langle T_{vv}\rangle
 = - \frac{c}{24 \pi} \{U,u \}= \frac{c}{48 \pi} \kappa^2.
 \label{2ndrank}
\end{equation}
If we put $c=1$ ($c=1/2$), this coincides with the energy flux $F^-_2$
($F_2^+$) 
from the thermal radiation of black holes.

The purpose of the present paper is to reproduce all the fluxes
$F_{2n}^{\pm}$ from conformal transformation properties 
of higher-spin currents. 

\section{Fourth-rank current} 

In order to reproduce the full spectrum of
thermal Hawking radiation, we need to calculate all the moments $F_{2n}$.  We
can obtain these higher moments by considering 
behavior of higher-spin currents in the black hole background.
First
let us consider a fourth-rank current for a scalar field $\phi$
\footnote{A third-rank conserved current does not exist for real
scalar fields.}.
 In flat $d=2$
space-time, there is a conserved traceless current
\begin{eqnarray}
J_{\mu \nu \rho \sigma}
&=& (8 \ \partial_{\mu} \phi \partial_\nu\partial_\rho\partial_{\sigma} \phi 
-12 \partial_\mu\partial_{\nu} \phi \partial_\rho\partial_{\sigma} \phi
 \nonumber    \\
 && -4  \ 
   g_{\mu \nu}\partial^\lambda \phi 
   \partial_\lambda\partial_\rho\partial_{\sigma} \phi
    + 8 \ 
   g_{\mu \nu}\partial^\lambda\partial_\rho \phi
   \partial_\lambda\partial_{\sigma} \phi
 \nonumber \\
 && -  
  g_{\mu\nu} g_{\rho\sigma}
  \partial^\lambda\partial^\tau \phi \partial_\lambda\partial_\tau \phi)
  + \mbox{symm.} 
\end{eqnarray}
The 'symm.' means symmetrization under $(\mu \nu \rho \sigma).$ 
Its holomorphic component is given by a sum of two terms
$J_{uuuu} \propto \partial_u \phi \partial_u^3 \phi -3/2 \ \partial_u^2 \phi
\partial_u^2 \phi.$ 
In curved space the current will acquire trace anomaly
and the holomorphy will be lost, but it is plausible to think that we can
separate the non-holomorphic part as we did in the case of energy-momentum
tensor and define a holomorphic fourth-rank current, whose conformal
transformation laws contain a generalization
 of the Schwarzian derivative.
{}From the OPE between $J_{uuuu}$ and $T_{uu}^{(conf)}$, we can obtain a
transformation law of $J_{uuuu}$ under an infinitesimal 
conformal transformation.  In
order to obtain its finite transformation law,
we follow the derivation of the ordinary Schwarzian derivative 
in \cite{yellow}. Two operators,
$:-\partial_u \phi \partial_u^3 \phi :$ and 
$:\partial_u^2 \phi \partial_u^2\phi:$, 
have different transformation properties but we can show  that
they give the same value of the Schwarzian derivatives for our specific
transformation from $u$ to the Kruskal $U$.  Therefore we will restrict
ourselves to consider only the first type 
$:-\partial \phi \partial^3\phi(u):$.  
The normal ordering of the operator is defined by a point splitting method as 
\begin{equation}
:\partial \phi \partial^3 \phi(u):
=\lim_{\epsilon\rightarrow 0} 
\left(\partial \phi(u+\epsilon/2) \partial^3 \phi(u-\epsilon/2) 
 +\frac{3}{2\pi\epsilon^4} \right).
\end{equation}
Here 
$\phi(u_1,v_1)\phi(u_2,v_2) \sim -\left(1/4\pi\right) \ln (u_1-u_2)(v_1-v_2)$.
Conformal transformations  of the r.h.s. under $u \rightarrow w(u)$
can be easily calculated. Taking the limit $\epsilon\rightarrow 0$, 
we obtain 
\begin{eqnarray}
 && :\partial \phi \partial^3 \phi(u): 
 \nonumber \\ 
 && = 
  w'w''': \partial \phi^{(w)} \partial\phi^{(w)}(w): 
  +3(w')^2w'' : \partial \phi^{(w)} \partial^2\phi^{(w)}(w): 
  \nonumber \\
 && +(w')^4: \partial \phi^{(w)}\partial^3\phi^{(w)}(w):  
  -\frac{1}{480\pi} \{w,u \}_{(1,3)}, 
  \label{tr13}
\end{eqnarray}
where
\begin{eqnarray}
 \{w,u \}_{(1,3)}&=& 6\frac{w^{'''''}}{w'} 
  -20\left( \frac{w'''}{w'} \right)^2 
  -45\left(\frac{w''}{w'}\right)^4
  \nonumber \\
  && 
  +90\frac{(w'')^2w'''}{(w')^3}  
    -30 \frac{w^{''''}w''}{(w')^2}
  \label{sd13}
\end{eqnarray}
is a generalized  Schwarzian derivative for the operator 
$:\partial \phi \partial^3 \phi(u):$.

We apply this transformation law to the conformal transformation from $u$ to
the Kruskal coordinate $w=U=-e^{-\kappa u}$.  By imposing the regularity
condition at the future horizon in the Kruskal coordinate, it is necessary
to require that all the operators in the Kruskal coordinate must finite at
the horizon.  Hence 
 the outgoing flux of the fourth rank current 
 from the black hole in the $(u,v)$
coordinate is given by
\begin{align} 
 \langle :-\partial \phi \partial^3 \phi(u): \rangle=
  \frac{1}{480\pi} \{ U, u \}_{(1,3)}=
 \frac{\kappa^4}{480\pi}. 
 \label{4thrank}
\end{align} 
This precisely agrees with the 3rd moment $F_4^-$ for bosons.

\section{Higher-spin currents for bosons}

The derivation of $F^-_4$ can be generalized to the 
higher moment $F^-_{2n}$ 
whose current is given by a linear combination of
$\partial_{\mu_1} \dots \partial_{\mu_m} \phi \ 
\partial_{\mu_{m+1}} \dots \partial_{\mu_{2n}} \phi$.
The holomorphic component is a linear combination of
$:(-1)^{n+m}\partial_u^m \phi \partial_u^{2n-m} \phi:$. 
Since we can show that all these terms give the same value of 
the Schwarzian derivative
for the conformal transformation $u \rightarrow U$, 
we consider the $m=1$ case.
A conformal transformation for 
$:(-1)^{n-1} \partial \phi \partial^{2n-1} \phi:$
can be more easily obtained by introducing its generating function
\begin{equation}
 :\partial  \phi(u) \partial \phi(u+a): \equiv
  \sum_{n=0}^\infty \frac{a^n}{n!} 
  :\partial \phi(u) \partial^{n+1} \phi(u):
\end{equation}
than calculating each term separately 
\footnote{Here we have summed over both of even and odd-rank currents
for notational simplicity. Since the odd-rank currents 
can be written as total derivatives,
their expectation values  vanish identically for a translationally invariant
system.}. 
Its conformal transformation under 
$u \rightarrow w(u)$
can be calculated  as
\begin{eqnarray}
&&  :\partial \phi(u) \partial\phi(u+a): \nonumber \\
 && = \partial_u w(u) \partial_u w(u+a)
  :\partial_w \phi^{(w)}(w(u)) \partial_w \phi^{(w)}(w(u+a)):
  \nonumber \\ &&
   + A_b(w,u),
   \end{eqnarray}
where $A_b(w,u)$ is a generating function of the generalized 
Schwarzian derivatives for bosons $\{ w, u\}_{(1, n)}$ and given by
\begin{equation}
 A_b(w,u) = -\frac{1}{4\pi}
  \frac{\partial_u w(u)\partial_u w(u+a)}{(w(u) - w(u+a))^2}
  +\frac{1}{4\pi a^2}.
\end{equation}
In our  case, $w=U= - e^{-\kappa u}$, 
the conformal transformation becomes
\begin{eqnarray}
&&
 :\partial_U \phi^{(U)}(U(u)) \partial_U \phi^{(U)}(U(u+a)):
 \nonumber \\ &&
  = e^{\kappa a} \left(\frac{1}{\kappa U}\right)^2  
  \left[
   :\partial \phi(u) \partial\phi(u+a): - A_b(U,u)
  \right],
\end{eqnarray}
and
\begin{eqnarray}
 A_b(U,u) = - \frac{\kappa^2}{16\pi}\frac{1}{\sinh^2 \frac{\kappa a}{2}}
  + \frac{1}{4\pi a^2}.
\end{eqnarray}
Regularity at the future horizon in the Kruskal coordinate determines the
expectation value of the generating function of the higher-rank currents as
\begin{equation}
\langle  
:\partial \phi(u) \partial \phi(u+a):
\rangle = A_b(U,u).
\end{equation}
$A_b(U,u)$ can be expanded as the following power series of $a$,
\begin{eqnarray}
 A_b (U, u)= \sum_{n=0}^\infty (-1)^n
  \frac{B_{n+1}\kappa^{2(n+1)}}{8\pi(n+1)}\frac{a^{2n}}{(2n)!}.
\end{eqnarray}
Hence the outgoing flux of the $2n$-th rank current for bosons 
is given by 
\begin{equation}
\langle :(-1)^{n-1}\partial \phi \partial^{2n-1} \phi(u): \rangle
=  \frac{1}{8\pi n} B_n\kappa^{2n}. 
\label{higherrank}
\end{equation}
This reproduces the flux from thermal Hawking radiation $F^-_{2n}$.
The relation (\ref{higherrank}) 
is a higher-spin generalization of (\ref{2ndrank}) and
(\ref{4thrank}).
We now conclude that the conformal field theory technique gives
the full thermal spectrum  of Hawking radiation for the bosonic case.

Two comments are in order.
First, by using the same technique of the generating function, we can 
show that the Schwarzian derivative for the operator 
$:(-1)^{n+m} \partial^m \phi \partial ^{2n-m} \phi:$ 
gives  the same value
 $ (B_n/8 \pi n) \kappa^{2n}$ for its expectation value
 in the black hole background. 
Or it is equivalent to  state that the  Schwarzian derivative 
vanishes for
a total derivative operator 
$\partial \left(  \partial^k \phi \partial ^l \phi \right)$ 
under the conformal transformation $u \rightarrow U$,
and we do not need to worry which linear combination of them
is a holomorphic component of the conserved current.

The second comment is the meaning of $A_b(U,u).$
It is a generating function of the 
Schwarzian derivatives of higher-spin currents 
under a conformal transformation $u \rightarrow U =-e^{\kappa u}$.
It can be written in an integral form as 
\begin{equation}
A_b (U,u) =  \int_0^\infty  \frac{d \omega}{2 \pi}
\omega  N^-(\omega) \cos(a \omega). 
\end{equation}
This is the temperature-dependent part of a finite temperature 
Green function for 
$\langle T \partial \phi(x) \partial \phi(x+a) \rangle_\beta$
as can be seen from
\begin{eqnarray}
 && \langle T \phi(x) \phi(y) \rangle_\beta \nonumber \\
 && = \int \frac{d^2 k}{(2 \pi)^2} 
  \left( 
   \frac{i}{k^2 + i\epsilon} + 2 \pi N^-(|\omega|) \delta(k^2) 
  \right) e^{-ik (x-y)}. \nonumber \\
\end{eqnarray}
It is the reason why 
we can reproduce the fluxes of all the higher-spin currents.
Appearance of the finite temperature Green function 
\cite{Gibbons-Perry} 
is natural
because the conformal transformation from $u$ to 
the Kruskal coordinate $U=-e^{-\kappa u}$ is nothing but
a conformal transformation from zero temperature to finite
temperature.

\section{Higher-spin currents for fermions}

The calculation can be similarly applied to a fermionic case.
A difference here is that  spinor has a conformal weight 1/2
and transforms as
$\psi(z) = \left(\partial_z w(z)\right)^{1/2} \psi^{(w)}(w(z))$.
In order to calculate the conformal transformation of operators like
$:\psi\partial^n \psi(z): $, we again consider its generating function;
\begin{equation}
 :\psi (z) \psi(z+a): 
  \equiv \sum_{n=0}^{\infty} \frac{a^n}{n!}:\psi\partial^n \psi(z):. 
\end{equation}
Under a conformal transformation, we have
\begin{eqnarray}
&&
 :\psi (u) \psi(u+a): 
  = \left( \partial_u w(u) \partial_u w(u + a) \right)^{1/2} 
  \nonumber \\ && \times
  :\psi^{(w)}(w(u))\psi^{(w)}(w(u+a)): 
  + A_f (w,u),
\end{eqnarray}
where
\begin{eqnarray}
 A_f (w,u)
  = \frac{i}{2\pi}
  \frac{\left(\partial_u w(u) \partial_u w(u + a) \right)^{1/2}}
         {w(u+a) - w(u)} 
  - \frac{i}{2 \pi a}.
\end{eqnarray}
Here we have used $\psi(u_1)\psi(u_2) \sim -i/2 \pi (u_1-u_2)$.
In the case, $w=U=-e^{-\kappa u}$, it becomes
\begin{eqnarray}
 A_f (U, u)= \frac{i}{2 \pi a}
  \left(\frac{\kappa a}{2} \frac{1}{\sinh\frac{\kappa a}{2}} - 1\right).
\end{eqnarray}
This is the temperature-dependent part of a finite temperature Green
function for fermions.
Expanding this as a power series of $a$, we find
\begin{eqnarray}
 A_f (U,u)= 
  \sum_{n=1}^{\infty} i^{2n+1}\frac{(1-2^{1-2n})B_n 
  \kappa^{2n}}{4 \pi n}
   \frac{a^{2n-1}}{(2n-1)!}.
\end{eqnarray}
Hence the flux of the $2n$-th rank current from a black hole
is given by the value of its Schwarzian derivative as 
\begin{eqnarray}
 \langle : \frac{i^{2n-1}}{2}\psi\partial^{2n-1} \psi(u): \rangle
  = \frac{\left(1-2^{1-2n}\right)B_n}{8 \pi n} \kappa^{2n}.
\end{eqnarray}
This again precisely gives the  thermal flux $F_{2n}^+$ for fermions.
The difference in the Planck distribution 
between scalars and spinors comes from the difference of conformal weights.
It is amusing to consider fields with another conformal weight to
derive possibly different thermal spectrum from $N^{\pm}(\omega)$.



\section{Discussion}

In this paper, we have applied a conformal field theory technique to obtain
the full thermal spectrum from Hawking radiation in both cases of bosons and
fermions.  We have shown that, 
by investigating
transformation properties of higher-spin currents under a conformal
transformation which maps the null coordinate $u$ to the Kruskal coordinate
$U=-e^{-\kappa u}$, 
the expectation value of each higher-spin current in the Unruh vacuum exactly
coincides with a corresponding specific
moment of the Planck distribution.
The full thermal distribution of the Hawking radiation
can be reproduced from these expectation values for the higher-spin currents.
In the previous analyses using anomalies in 
 \cite{Robinson:2005pd,IUW1,IUW2,Murata,Das, setare, Xu, IMU, Jiang, 
Jiang2,CF},  only the zeroth and first moments of the Planck
distribution, corresponding to fluxes of charge (or angular momentum) and
energy, could be derived. 
In this sense, our present calculation has made the previous partial results
complete. 
But it is  more desirable if we can reinforce the present
conformal field theory calculation by a covariant 
calculation of higher-spin currents.
It is known that even in the gravitational and gauge field
background, two-dimensional massless field theories 
can be described by conformal field theories,
and the present calculation of using
holomorphic (or anti-holomorphic) higher-spin
currents is naturally justified.
In a covariant formulation, 
these higher-spin currents  
are given by  specific components of covariant
tensors.
To clarify them, it is necessary first to  construct 
covariant traceless higher-spin tensors
and then to calculate
the trace anomalies of these tensors.
By solving these equations, 
we should be able to extract the (anti-)holomorphic components,
corresponding to those we have used in our paper.
It is also interesting to see what kind of 
transformations these higher-spin currents will generate as the conformal
energy-momentum tensor generates conformal transformations
which are combinations of coordinate transformations and Weyl
transformations. We will come back to these problems 
in near future \cite{IMU2}.

\begin{acknowledgments}
We would like to thank H. Itoyama for his fruitful comments.
T. M. was supported in part by a JSPS Research Fellowship for Young
Scientists. 
\end{acknowledgments}


\begin{thebibliography}{99}

\bibitem{Hawking:sw}
S.~Hawking,
Commun.\ Math.\ Phys.\  {\bf 43}, 199 (1975).

\bibitem{Hawking:rv}
S.~Hawking,
Nature (London) {\bf 248}, 30 (1974).

\bibitem{Unruh:db}
W.~Unruh,
Phys.\ Rev.\ D {\bf 14}, 870 (1976).

\bibitem{Robinson:2005pd}
S.~P.~Robinson and F.~Wilczek,
Phys.\ Rev.\ Lett.\  {\bf 95}, 011303 (2005).
  
\bibitem{IUW1}
S.~Iso, H.~Umetsu and F.~Wilczek,
Phys.\ Rev.\ Lett.\  {\bf 96}, 151302 (2006).
  
\bibitem{IUW2}
S.~Iso, H.~Umetsu and F.~Wilczek,
Phys.\ Rev.\ D {\bf 74}, 044017 (2006).

\bibitem{Murata} 
K.~Murata and J.~Soda,
Phys.\ Rev.\ D {\bf 74}, 044018 (2006).
 
\bibitem{Das} 
E.~Vagenas and S.~Das,
JHEP {\bf 0610}, 025 (2006).

\bibitem{setare} 
M.~R.~Setare,
arXiv:hep-th/0608080.

\bibitem{Xu} 
Z.~Xu and B.~Chen, 
arXiv: hep-th/0612261.


\bibitem{IMU}
 S.~Iso, T.~Morita and H.~Umetsu,
 arXiv:hep-th/0612286.
  
\bibitem{Jiang}
Q.~Q.~Jiang and S.~Q.~Wu,
arXiv:hep-th/0701002.
 
\bibitem{Jiang2}
Q.~Q.~Jiang, S.~Q.~Wu and X.~Cai,
arXiv:hep-th/0701048, hep-th/0701235.

\bibitem{IMU2}
S.~Iso, T.~Morita and H.~Umetsu,
under investigations

\bibitem{Thorlacius}
L.~Thorlacius,
Nucl.\ Phys.\ Proc.\ Suppl.\  {\bf 41}, 245 (1995).

\bibitem{Strm}
A.~Strominger,
arXiv:hep-th/9501071.
 
  
\bibitem{CF}
S.~Christensen and S.~Fulling,
Phys.\ Rev.\ D {\bf 15}, 2088 (1977).
  
\bibitem{yellow}
P.~Di Francesco, P.~Mathieu and D.~Senechal, {\it Conformal Field Theory}
(Springer-Verlag, 1997).

\bibitem{Gibbons-Perry}
G.~W.~Gibbons and M.~J.~Perry,
Proc.\ Roy.\ Soc.\ Lond.\ A {\bf 358}, 467 (1978).


\end{thebibliography}
\end{document}